\documentclass[12pt]{iopart}
\usepackage{color}
\usepackage{graphicx}
\usepackage{amssymb}

\begin{document}

\title{Conservation rules for entanglement transfer between qubits}

\author{Stanley Chan$^1$, M D Reid$^2$ and Z  Ficek$^3$}

\address{$^1$Department of Physics, School of Physical Sciences, The University of Queensland, Brisbane, QLD 4072, Australia}
\address{$^2$ARC Centre of Excellence for Quantum-Atom Optics and Centre for Atom
Optics and Ultrafast Spectroscopy, Swinburne University of Technology, Melbourne, Australia}
\address{$^3$The National Centre for Mathematics and Physics, KACST, P.O. Box 6086, Riyadh 11442, Saudi Arabia}
\eads{\mailto{zficek@kacst.edu.sa}}

\begin{abstract}
We consider an entangled but non-interacting qubit pair $a_{1}$ and $b_{1}$ that are independently coupled to a set of local qubit systems, $a_{I}$ and $b_{J}$, of $0$-bit value,
respectively. We derive rules for the transfer of entanglement from the pair $a_{1}-b_{1}$ to an arbitrary pair $a_{I}-b_{J}$, for the case of qubit-number conserving local interactions. It is shown that the transfer rule depends strongly on the initial entangled state. If the initial entanglement is in the form of the Bell state corresponding to anti-correlated qubits, the sum of the square of the non-local pairwise concurrences is conserved. If the initial state is the Bell state with correlated qubits, this sum can be reduced, even to zero in some cases, to reveal a complete and abrupt loss of all non-local pairwise entanglement. We also identify that for the nonlocal bipartitions $A-b_{J}$ involving all qubits at one location, with one qubit $b_{J}$ at the other location, the concurrences satisfies a simple addition rule for both cases of the Bell states, that the sum of the square of the nonlocal concurrences is conserved.
\end{abstract}


\submitto{\JPB}

\maketitle

\section{Introduction}

Entanglement is not only crucial to the transition between classical and quantum behavior~\cite{Bell}, but is essential to many key applications in quantum information~\cite{genqinf}. An understanding of how entanglement is transferred between systems and the existence of associated conservation rules is of fundamental and practical importance. 

The seminal works of Bell focused on entanglement that is shared between
two distant and non-interacting {}``qubit'' (two-level) systems~\cite{zeilrmp}. 
This {}``nonlocal'' entanglement can be maintained
over long distances with exciting implications for tests of quantum
mechanics and applications such as quantum cryptography~\cite{crplong}.
However, a fundamental issue is the degradation of entanglement brought
about because each party inevitably interacts {}``locally'' with
other systems. This local coupling can lead to an abrupt
depletion of the original entanglement~\cite{yueberly,d03,dh04,ja06,ao08}, an effect
which has been recently experimentally confirmed~\cite{alexp}. Under some circumstances, the already lost entanglement can revival after a finite time~\cite{ft06,cz08,ls10,xf10}. The
dynamical behavior of entanglement under the action of the environment
is regarded as a central issue in quantum information~\cite{ebscience,lopez}. 

Intuition tells us that the two-qubit entanglement is not truly {}``lost'', but simply redistributed among the interacting parties~\cite{lopez,yonac,yonac1,sainz,stan}.
While it is the case that entanglement can be created between local
systems, due to the local couplings, it is logical to investigate
under what circumstances a global nonlocal entanglement is conserved,
to reflect that no further {}``nonlocal'' interaction has taken
place, and to ask whether a rule exists to describe the entanglement
transfer and to express a conserved global entanglement in terms of
the constituent entanglement.

Indeed a requirement of a measure of entanglement between remotely
separated parties is that the entanglement does not increase under
certain local operations assisted by classical communication~\cite{entmeasLU}.
Our interest here is more specific. We construct subsets of local
systems, and consider only qubit-conserving interactions between them,
so a transfer of qubits is described. We do not allow further interaction
between the two remotely separated systems themselves. 

The question of whether a universal conservation rule exists to describe
the entanglement transfer is a difficult one in full, requiring knowledge
of necessary and sufficient entanglement measures where entanglement
can be shared among more than two qubits. For example, the recent
work of Hiesmayr \etal~\cite{hiesmulti} focuses on the development
of computable measures of entanglement in the multipartite scenario.
Nonetheless, in this paper we take a first step by deriving some simple
conservation rules that apply for the fundamental case of an initial
two qubit {}``Bell state'' entanglement. The rules allow a quantitative
knowledge of the degree of the {}``nonlocal'' entanglement, in terms
of the concurrence measure~\cite{wit2}, shared between remaining
parties, given a knowledge of the entanglement between one spatially
separated pair.

\begin{figure}
\begin{center}
\includegraphics[width=0.5\columnwidth]{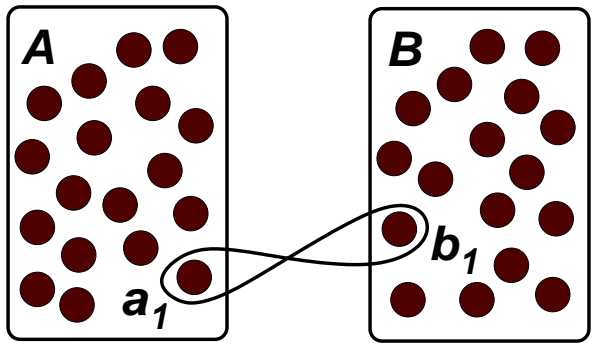}
\end{center}
\caption{We derive rules to describe how the entanglement is transferred among
qubits, when each member of an entangled non-interacting qubit pair $a_{1}$, $b_{1}$ is locally and independently coupled to a set of qubits, $a_{I}$ and $b_{I}$, respectively.}
\label{fig1}
\end{figure}

Specifically, we examine entanglement transfer from one qubit pair $a_{1},b_{1}$ into a set of qubits pairs $a_{I},b_{J}$, where members of all {}``nonlocal'' pairs $a_{I},b_{J}$ are non-interacting, as illustrated in figure~\ref{fig1}. The initial entanglement is in the form of a Bell state, with all other qubits having $0$ bit value, and the local interaction that can exist within the qubit sets, $A\equiv\left\{ a_{I}\right\}$ and $B\equiv\left\{ b_{J}\right\}$. The interaction is constrained only by the requirement that there is a transfer of qubit value, so the total local qubit number, or Hamming weight, is conserved. Such interactions are relevant to quantum networks, in creating qubit superposition states, and are fundamental in describing system-environment interactions~\cite{yueberly,qczoll}.

Yonac \etal~\cite{yonac,yonac1} have presented a conservation rule
for the case of Bell states with anti-correlated qubits
\begin{equation}
|\Psi\rangle=\cos\alpha|10\rangle+\sin\alpha\, {\rm e}^{i\beta}|01\rangle ,\label{eq:bell1}
\end{equation}
where $\beta$ is an arbitrary phase and $|10\rangle$ represents qubit values of $1$ and $0$ for the qubit pair $a_{1}$ and $b_{1}$ respectively. Yonac \etal observed that $C_{11}+C_{22}$ is constant, where $C_{IJ}$ is the concurrence measure~\cite{wit2} of entanglement between qubits $a_{I}$ and $b_{J}$. Their result however was specific to one local Hamiltonian, that of the Jaynes-Cummings model, and is not valid generally. 

Lopez \etal~\cite{lopez} presented a very different result for Bell states with correlated qubits
\begin{equation}
|\Phi\rangle=\cos\alpha|11\rangle+\sin\alpha\, {\rm e}^{i\beta}|00\rangle .\label{eq:bell2}
\end{equation}
They showed that when such a qubit pair is
coupled to independent reservoirs, the death of qubit entanglement
precedes the birth of reservoir entanglement, so there is a temporary
loss of all nonlocal pairwise entanglement. Again, their results applied
to one form of local interaction Hamiltonian only. 

Our conservation rules apply to all cases of qubit-preserving {}``local''
interactions, and allow new insight. Which conservation rule is valid
depends only on the type of global entanglement that is imprinted
onto the system. When the initial {}``global'' entanglement is that
of the Bell state $|\Psi\rangle$, a remarkably simple conservation
result exists
\begin{equation}
\sum_{I=1}^{N}\sum_{J=1}^{M}C_{IJ}^{2}(t)=C_{AB}^{2} , \label{eq:concurrenceequal}
\end{equation}
which shows that the total nonlocal pairwise entanglement is conserved. However, this conservation rule manifests in the {\it square} of the concurrence (the {}``tangle''), not in the
concurrence itself. The conserved quantity corresponds to the initial
entanglement, but only when evaluated as $C_{AB}^{2}$, the square
of the concurrence $C_{AB}$ of the Bell state. For this case, we
will prove an additivity of constituent entanglement, that the entanglement
shared between any two nonlocal partitions is the sum of the nonlocal
pairwise entanglement of the constituents of the partitions. This
rule applies to closed systems, and is investigated for open systems. 

When the initial {}``global'' entanglement is that of the Bell state $|\Phi\rangle$, the following inequality holds
\begin{equation}
0\leq\sum_{I,J}C_{IJ}^{2}(t)\leq C_{AB}^{2} .\label{eq:concurrenceunequal}
\end{equation}
We will show there can be a vanishing of all nonlocal pairwise
constituent entanglement $C_{IJ}$, despite conservation of global
nonlocal entanglement $C_{AB}$, to give consistency with the result
of Lopez \etal~\cite{lopez}.

The two different scenarios may be thought of in the following way.
Suppose entanglement exists between $A$ and $B$, where $A$ is made
up of subsystems measured by Alice, Ann, Agatha respectively, and
$B$ is composed of subsystems measured by Bob, Bill and Brian. In
the $\Psi$ scenario, the global nonlocal entanglement can \emph{alway}s
be evaluated, through measurements shared only by pairs: Alice-Bob,
Alice-Bill, Ann-Bob, Ann-Bill, etc. In the $\Phi$ scenario, this
is not the case. Examples exist where all pairwise entanglement would
be zero, despite there being a global entanglement, between $A$ and
$B$. This has potential implications for quantum cryptography, where
measurement of shared entanglement between two parties, $A$ and $B$,
at different locations is used to determine security~\cite{bellcry}. 

We will show however, that in both cases, the global (original) entanglement
could be inferred with communication between all parties at $A$ and
one at $B$: Alice-Ann-Agatha-Bob; Alice-Ann-Agatha-Bill, and so on.
This follows from the simple addition rule that can be proved in both
cases 
\begin{equation}
C_{AB}^{2}=C_{AB_{1}}^{2}+C_{AB_{2}}^{2}+\ldots +C_{AB_{N}}^{2} .\label{eq:sumonepart}
\end{equation}
This represents saturation of the CKW inequality~\cite{3tangle},
which constrains the concurrences for any three parties $A$, $B_{1}$,
$B_{2}$, according to
\begin{equation}
C_{AB_{1}}^{2}+C_{AB_{2}}^{2}+\ldots +C_{AB_{N}}^{2}\leq C_{AB}^{2} .\label{eq:ckw}
\end{equation}
The equality (\ref{eq:sumonepart}) is useful to cryptography scenarios,
where $B$ and $A$ share an entangled Bell state, but a third party
Eve (who we call $B_{2}$) may eavesdrop through an interaction that
conserves qubit number. In this case, regardless of the type of Bell
state used to transport the entanglement, Bob (who we call $B_{1}$)
and Alice (who we call $A$) can deduce the degree of entanglement
that Eve can possess ($C_{AB_{2}}^{2}=C_{AB}^{2}-C_{AB_{1}}^{2}$).

Studies of entanglement distribution among $m$ parties have revealed
that for some states (GHZ-type states) entanglement can exist among
$m$ parties, but not exist when measurements are performed on less
than $m$ parties. On the other hand, for other states, the $W$-states,
distributed entanglement satisfies a simple rule based on saturation
of the CKW inequality. The rules describing entanglement transfer
are largely determined by what states can be formed under the local
entanglement transfer interactions.

\section{Qubit-conserving Local Interaction Hamiltonian Model}

We begin by considering the two global non-interacting systems, the
qubit sets, $A\equiv\left\{ a_{I}\right\}$ and $B\equiv\left\{ b_{J}\right\}$.
For the initial states we consider, the total qubit number (or Hamming
weight) $Q_{A/B}$ of $A/B$ has a value $0$ or $1$. We thus define
eigenstates $|Q_{A}\rangle|Q_{B}\rangle$, where $Q_{A},Q_{B}=0,1$
are the outcomes for the total qubit number at $A$ and $B$, respectively.
We assume each system $A$, $B$ has constituent subsystems, and there
are internal interactions between them, described by Hamiltonians
$H_{A}$ and $H_{B}$ respectively, so that the total Hamiltonian is 
\begin{equation}
H=H_{A}+H_{B} .\label{eq:ham}
\end{equation}
Importantly, it is assumed in this model that each of $H_{A}$ and $H_{B}$ conserves the qubit number (Hamming weight) of the system $A$ and $B$, respectively. 

A measure of bipartite entanglement between two qubit systems, such as $A$ and $B$, is the concurrence, defined as $C_{AB}={\rm \max}\left(0,\sqrt{\lambda_{1}}-\sqrt{\lambda_{2}}-\sqrt{\lambda_{3}}-\sqrt{\lambda_{4}}\right)$, where $\lambda_{1,}\ldots,\lambda_{4}$ are the eigenvalues, in decreasing order, of the density matrix $\rho'=\rho(\sigma_{y}\otimes\sigma_{y})\rho^{\ast}(\sigma_{y}\otimes\sigma_{y})$. Here $\rho$ is the two-qubit system density operator, $\rho^{*}$ is the complex conjugation of $\rho$ in the standard basis, and $\sigma_{y}$ is the Pauli matrix in the $y$ direction expressed in the same basis. The maximum possible entanglement is given by $C_{AB}=1$, while $C_{AB}=0$ implies no entanglement. The state is entangled when $C_{AB}>0$. 

The global bipartite entanglement defined for systems $A$ and $B$ is invariant throughout the evolution. This follows from the fact that the Hamiltonian (\ref{eq:ham}) does not allow interaction between the two systems, or with external systems.

It is useful to first examine the full qubit description of the system,
which does evolve. Since the total qubit value at each of $A$ and
$B$ is assumed conserved under action of the Hamiltonian~(\ref{eq:ham}) and
we consider the case where initially only $a_{1}$ and $b_{1}$ can
have a nonzero qubit value, we then see that at any time there can be at
most one qubit system at each of $A$ and $B$ with a bit value 1,
all other bits are 0. The possible states of the system can thus be
written as
\begin{equation}
|\left\{ 1\right\} _{n}\rangle_{A}|\left\{ 1\right\} _{m}\rangle_{B} ,\label{eq:state11}
\end{equation}
where $|\left\{ 1\right\} _{n}\rangle_{A}=|0\ldots\left\{ 1\right\} _{n}\ldots0\rangle$
denotes that the qubit $a_{n}$ is in state $|1\rangle$ while all
other qubits $a_{I\neq n}$ are in $|0\rangle$, and $|\left\{1\right\}_{m}\rangle_{B}$
similarly defines the qubit states of $B$. Also possible are the
states, written as $|0\rangle_{A}$ and $|0\rangle_{B}$, with zero
qubit value, for all qubits. Thus, explicitly, if we write the initial
state as a superposition of the two Bell states
\begin{eqnarray}
|\psi(0)\rangle & = & d_{00}(0)|0\rangle_{A}|0\rangle_{B}+d_{01}(0)|0\rangle_{A}|\{1\}_{1}\rangle_{B}\nonumber\\
&& +d_{10}(0)|\{1\}_{1}\rangle_{A}|0\rangle_{B}+d_{11}(0)|\{1\}_{1}\rangle_{A}|\{1\}_{1}\rangle_{B} ,\label{eq:initial state10}
\end{eqnarray}
then the evolution is given by 
\begin{eqnarray}
\fl \psi(t)\rangle = {\rm e}^{-iHt/\hbar}|\psi(0)\rangle
= d_{00}(0)|0\rangle_{A}|0\rangle_{B}+d_{01}(0)|0\rangle_{A}{\rm e}^{-iH_{B}t/\hbar}|\{1\}_{1}\rangle_{B}\nonumber \\
+ d_{10}(0){\rm e}^{-iH_{A}t/\hbar}|\{1\}_{1}\rangle_{A}|0\rangle_{B}
+d_{11}(0){\rm e}^{-iH_{A}t/\hbar}|\{1\}_{1}\rangle_{A}{\rm e}^{-iH_{B}t/\hbar}|\{1\}_{1}\rangle_{B}\nonumber \\
= d_{00}(0)|0\rangle_{A}|0\rangle_{B}+\sum_{m=1}^{M}d_{0m}(t)|0\rangle_{A}|\left\{ 1\right\} _{m}\rangle_{B}\nonumber \\
+\sum_{n=1}^{N}d_{n0}(t)|\left\{ 1\right\} _{n}\rangle_{A}|0\rangle_{B} 
+\sum_{n=1}^{N}\sum_{m=1}^{M}d_{nm}(t)|\left\{ 1\right\} _{n}\rangle|\left\{ 1\right\} _{m}\rangle ,\label{eq:evolution}
\end{eqnarray}
where 
\begin{eqnarray}
d_{00}(t) &=& d_{00}(0) ,\quad |d_{10}(0)|^{2} = \sum_{n=1}^{N}|d_{n0}(t)|^{2} ,\nonumber\\
|d_{01}(0)|^{2} &=& \sum_{m=1}^{M}|d_{0m}(t)|^{2} ,\quad 
|d_{11}(0)|^{2} = \sum_{n=1}^{N}\sum_{m=1}^{M}|d_{nm}(t)|^{2} . \label{eq:probconserv}
\end{eqnarray}
In the derivation of (\ref{eq:evolution}) we have used the Hamiltonian (\ref{eq:ham}), that $H=H_{A}+H_{B}$, and that the unitary operators conserve probability, so for example 
\begin{eqnarray}
{\rm e}^{-iH_{B}t/\hbar}|\{1\}_{1}\rangle_{B} = \sum_{n=1}^{N}c_{n}(t)|\{1\}_{n}\rangle_{B} ,\label{eq:probconser2}
\end{eqnarray}
where $\sum_{n=1}^{N}|c_{n}(t)|^{2}=1$. Note that the coefficient $d_{00}$
is time-independent, because a system in the ground state remains
in the ground state under the action of the Hamiltonian. Moreover, the global bipartite entanglement between systems $A$ and $B$ remains
constant. This is because for evaluation of the bipartite concurrence
$C_{AB}$, the expansion of $|\psi(t)\rangle$ in terms of the eigenstates
$|Q_{A}\rangle|Q_{B}\rangle$ of the \emph{total }qubit values $Q_{A}$
and $Q_{B}$ at $A$ and $B$ is all that is required. The factorization
$\exp(-iHt/\hbar) = \exp(-iH_{A}t/\hbar)\exp(-iH_{B}t/\hbar)$ of the unitary
operator leads to the invariance of $C_{AB}$, since $H_{A}$, $H_{B}$
each conserve the total bit values of $A$ and $B$, respectively.
Hence, the concurrence $C_{AB}$ is invariant.

\section{Entanglement evolution and conservation rules for Bell states}

We now examine the two types of pure state entanglement that can characterize
the two qubit system $A-B$~\cite{yueberly}. These are given by the
two Bell states, one with anticorrelated {}``spins'' 
\begin{equation}
|\Psi\rangle=\cos\alpha|1\rangle|0\rangle+{\rm e}^{i\beta}\sin\alpha|0\rangle|1\rangle\label{eq:entcons1}\end{equation}
and the other with correlated spins
\begin{equation}
|\Phi\rangle=\cos\alpha|1\rangle|1\rangle+{\rm e}^{i\beta}\sin\alpha|0\rangle|0\rangle.\label{eq:entcons2}\end{equation}
In both cases the concurrence is $C_{AB}=2\sin\alpha\cos\alpha$. It indicates that the maximal entanglement occurs at $\alpha=\pi/4$.

Next we examine what happens when each of the systems $A$ and $B$ is composed of a collection of \emph{interacting} qubits, which we denote $\left\{ a_{I}\right\} $ and $\left\{ b_{J}\right\}$, respectively. Since the overall qubit value at each of $A$ and $B$ must be conserved under the Hamiltonian~(\ref{eq:ham}), for the initial states $|\Psi\rangle$ and $|\Phi\rangle$, there can be at most one qubit at each of $A$ or $B$ with a bit $1$ and all other bits with $0$. The possible states for $t>0$ are described by~(\ref{eq:evolution}).

\subsection{Conservation rule for Bell state $|\Psi\rangle$}

We assume first that the initial state is the Bell state $|\Psi\rangle$.
Using the result (\ref{eq:evolution}) with $d_{00}(0)=d_{11}(0)=0$,
we find that the evolution for this form of global entanglement $|\Psi\rangle$
can be written
\begin{equation}
|\Psi(t)\rangle=\sum_{I=1}^{N}d_{AI}(t)|\left\{ 1\right\} _{I}\rangle|0\rangle+\sum_{J=1}^{M}d_{BJ}(t)|0\rangle|\left\{ 1\right\} _{J}\rangle .\label{eq:eqnbell1}
\end{equation}
Here, $d_{AI}(t)$ and $d_{BJ}(t)$ satisfy, after using (\ref{eq:probconserv}):
\begin{equation}
\sum_{I=1}^{N}|d_{AI}(t)|^{2}=\cos^{2}\alpha ,\quad \sum_{J=1}^{M}|d_{BJ}(t)|^{2}=\sin^{2}\alpha .\label{eq:prob}
\end{equation}
For $N=M=2$ this state can be written as 
\begin{equation}
|\Psi(t)\rangle=d_{A1}|1000\rangle+d_{A2}|0100\rangle+d_{B1}|0010\rangle+d_{B2}|0001\rangle ,\label{eq:eqn3explicit}
\end{equation}
where $|1000\rangle$ denotes that the qubit $a_{1}$ has value $1$,
while others are $0.$.., etc. This conservation of probability holds
because there is no external coupling, nor coupling between the systems
$A$ and $B$, that allows a transfer of qubit excitation~\cite{sainz}. 

We now relate the global bipartite entanglement $C_{AB}$, which is
conserved, to the non-local pairwise concurrences $C_{IJ}$ of subsystems
$a_{I}$ and $b_{J}$. This pairwise concurrence is calculated by
tracing over all other systems. Calculation shows the reduced density
matrix for the $a_{I},b_{J}$ system, written in the basis states $|11\rangle,|10\rangle$,
$|01\rangle$ and $|00\rangle$, is of {}``$X$-state'' form~\cite{yonac}: 
\begin{equation}
\rho_{IJ}=\left(\begin{array}{cccc}
0 & 0 & 0 & 0\\
0 & |d_{AI}|^{2} & d_{AI}d_{BJ}^{*} & 0\\
0 & d_{AI}^{*}d_{BJ} & |d_{BJ}|^{2} & 0\\
0 & 0 & 0 & \sum_{i\neq I}|d_{Ai}|^{2}+\sum_{j\neq J}|d_{Bj}|^{2}\end{array}\right) .\label{eq:matrixrho}
\end{equation}
In this case, the pairwise concurrence is simply given by 
\begin{equation}
C_{IJ} = 2|d_{AI}||d_{BJ}| .\label{eq:concdd}
\end{equation}

\noindent \textbf{Theorem 1:} For the case of a global Bell entanglement $C_{AB}$ of type $|\Psi\rangle$ between two non-interacting systems $A$ and $B$, the sum of the square of the pairwise constituent {}``nonlocal'' concurrences (SSPC) is conserved: SSPC$=\sum_{I=1}^{N}\sum_{J=1}^{M}C_{IJ}^{2}=C_{AB}^{2}.$
The entanglement shared by \emph{any} two nonlocal partitions $\left\{ a_{i},a_{j},\ldots\right\}$ and $\left\{ b_{m},b_{n},\ldots\right\}$ satisfies a simple Pythagorean addition of constituent
entanglement 
\begin{equation}
C_{AB}^{2}=C_{\left\{ i,j,\ldots\right\} \left\{m,n,\ldots\right\}}^{2} = \sum_{k=i,j,\ldots}\sum_{l=m,n,\ldots}C_{kl}^{2} .\label{eq:add}
\end{equation}

In addition, we can write the sum rule for the entanglement shared between system~$A$ and each of the subsystems of $B$: 
\begin{equation}
C_{AB}^{2}=C_{A\{m,n,\ldots\}}^{2}=\sum_{l=m,n,\ldots}C_{Al}^{2} .\label{eq:rulesum}
\end{equation}
\textbf{Proof}: For any Hamiltonian of the form $H=H_{A}+H_{B}$, the probability sums~(\ref{eq:prob}) are constant~\cite{sainz}. Hence, given that the pairwise concurrence is derived from~(\ref{eq:matrixrho}), the conjecture must hold
\begin{eqnarray}
{\rm SSPC} &=& \sum_{I}\sum_{J}|C_{IJ}|^{2}=4\left(\sum_{I=1}^{N}|d_{AI}(t)|^{2}\right)\times\left(\sum_{J=1}^{M}|d_{BJ}(t)|^{2}\right) \nonumber\\
&=& 4\cos^{2}\alpha\sin^{2}\alpha = C_{AB}^{2} .
\end{eqnarray}
The result (\ref{eq:rulesum}) follows in the same manner from direct evaluation of concurrences after tracing. We note other sum rules follow for this system. There is an additivity of constituent
entanglement, that the entanglement shared between any two nonlocal partitions is the sum of the nonlocal pairwise entanglement of the constituents of the partitions.

We note that the state (\ref{eq:eqnbell1}) reduces to the $2N$-partite
$W$-state $(|100\ldots\rangle+|010\ldots\rangle+|001\ldots\rangle+\ldots)/\sqrt{2N}$
when the probability amplitudes are equal. We note that if we consider only the entanglement between system $A$, which is written $A\equiv\left\{ i,j,\ldots\right\} $, and the subsystems of $B$, which we write $B_{1}, B_{2},\ldots$, then (\ref{eq:rulesum}) reduces to 
\begin{equation}
C_{AB}^{2}=C_{AB_{1}}^{2}+C_{AB_{2}}^{2}+\ldots +C_{AB_{N}}^{2} ,\label{eq:rulesum2}
\end{equation}
which is the well-known monogamy relation for $W$-states, for which the CKW inequality is saturated~\cite{3tangle,kd09}. The $W$-states \cite{vidthreequbit} are known to be robust with respect to particle losses, and this is reflected in the conservation rule, which states that entanglement
is preserved, with concurrence $C_{IJ}=1/N$, after tracing over all but two parties $I,J$-that is, if all but two parties lose the qubit information. That the $W$- state has the greatest amount
of pairwise bipartite entanglement possible after tracing over all other parties was conjectured by Dur \etal~\cite{vidthreequbit}. If we consider the three sub-systems $A,B_{1},B_{2}$, then the statement~(\ref{eq:rulesum2}) reduces to the result that the $3$-tangle defined
as $C_{A(B_{1}B_{2})}^{2}-C_{AB_{1}}^{2}-C_{AB_{2}}^{2}$ is $0$, which is a known result for the $4$-qubit $W$-state~\cite{fourqubit}.

\subsection{Concurrence inequality rule for Bell state $|\Phi\rangle$}

Where the global entanglement of the qubits $a_{1}, b_{1}$ is that of the Bell state $|\Phi\rangle$, the Pythagorean sum of the pairwise concurrences is no longer conserved, but satisfies
the inequality~(\ref{eq:concurrenceunequal}). In this case, the wave function evolves as 
\begin{equation}
|\Phi(t)\rangle=\sum_{i=1}^{N}\sum_{j=1}^{M}c_{ij}(t)|\left\{ 1\right\} _{i}\rangle|\left\{ 1\right\} _{j}\rangle+c_{0}|0\rangle|0\rangle .\label{eq:belleqn2}
\end{equation}
Here
\begin{equation}
|c_{0}(t)|^{2}=\sin^{2}\alpha ,\quad \sum_{i,j}|c_{ij}(t)|^{2} = \cos^{2}\alpha  \label{eq:prob2}
\end{equation}
as follows from (\ref{eq:probconserv}). For $N=2$, the state can be expressed as
\begin{eqnarray}
|\Phi(t)\rangle &=& c_{11}|1010\rangle+c_{12}|1001\rangle+c_{21}|0110\rangle+c_{22}|0101\rangle\nonumber \\
&& +c_{0}|0000\rangle .\label{eq:matrixstate2}
\end{eqnarray}
In this case, the reduced density matrix $\rho_{IJ}$ for the qubits $a_{I}$ and $b_{J}$ written in terms of basis states $|11\rangle$, $|10\rangle$, $|01\rangle$ and $|00\rangle$ is
\begin{equation}
\fl \rho_{IJ} = \left(\begin{array}{cccc}
|c_{IJ}|^{2} & 0 & 0 & c_{IJ}c_{0}^{*}\\
0 & \sum_{n\neq J}|c_{In}|^{2} & 0 & 0\\
0 & 0 & \sum_{m\neq I}|c_{mJ}|^{2} & 0\\
c_{IJ}^{*}c_{0} & 0 & 0 & \sum_{n,m\neq J,I}|c_{m,n}|^{2}+|c_{0}|^{2}
\end{array}\right). \label{eq:matrix2}
\end{equation}
For example, for the case of $N=M=2$, the reduced density matrix
for qubits~$a_{1}$ and~$b_{1}$ is
\begin{eqnarray}
\rho_{red} & = & \left( c_{0}|00\rangle+c_{11}|11\rangle\right)\left(c_{0}^{*}\langle00|+c_{1}^{*}\langle11|\right) \nonumber \\
 &  & +|c_{12}|^{2}|10\rangle\langle10|+|c_{21}|^{2}|01\rangle\langle01|
 +|c_{22}|^{2}|00\rangle\langle00| .\label{eq:fourqubitstate}
 \end{eqnarray}
The concurrence of (\ref{eq:matrix2}) is
\begin{equation}
\fl C_{IJ} = \max\left\{0, 2\left(|c_{IJ}||c_{0}|- \sqrt{\left(\sum_{n\neq J}|c_{In}|^{2}\right)\left(\sum_{m\neq I}|c_{mJ}|^{2}\right)}\right)\right\} ,\label{eq:conc}
\end{equation}
from which we note immediately that $C_{IJ}^{2}\leq4|c_{0}|^{2}|c_{IJ}|^{2}$
and thus, using (\ref{eq:prob2}), the inequality~(\ref{eq:concurrenceunequal})
must hold. Hence, we write the following theorem.\\

\noindent \textbf{Theorem 2:} For the case of a global Bell entanglement $C_{AB}$
of type $|\Phi\rangle$ between two non-interacting systems $A$ and $B$, the sum of the square of the pairwise constituent {}``nonlocal'' concurrences (SSPC) is constrained by an upper bound. The entanglement shared by \emph{any} two nonlocal partitions $\left\{ a_{i},a_{j},\ldots\right\}$, $\left\{ b_{m},b_{n},\ldots\right\} $ satisfies 
\begin{equation}
C_{AB}^{2}=C_{\left\{ i,j,\ldots\right\} \left\{ m,n,\ldots\right\} }^{2}\leq\sum_{k=i,j,\ldots}\sum_{l=m,n,\ldots}C_{kl}^{2} .\label{eq:ineq2}
\end{equation}
It is possible for \emph{all} pairwise entanglement to vanish. However, we find the following sum rule does hold, to give a saturation of the monogamy relation \cite{3tangle} for the total system at $A$, with the components of $B$: 
\begin{equation}
C_{AB}^{2} = C_{A\{m,n,\ldots\}}^{2}=\sum_{l=m,n,\ldots}C_{Al}^{2} .\label{eq:addipsi}
\end{equation}

\noindent {\bf Proof:} The proof of the inequality follows from above. In fact, the inequality~(\ref{eq:ineq2}) can be proved for any state via the CKW inequality~(\ref{eq:ckw}). One merely applies the CKW inequality a second time to each of the terms on the left side of~(\ref{eq:ckw}).

To prove the equality (\ref{eq:addipsi}), we note that the state (\ref{eq:belleqn2}) written in terms of the total qubit system at $A$ is $\sum_{j=1}^{M}\tilde{c}_{1j}|1\rangle|\left\{ 1\right\}_{j}\rangle+c_{0}|0\rangle|0\rangle$, which can be written explicitly as
\begin{eqnarray}
|\Phi(t)\rangle &=& \tilde{c}_{1}|1;100\ldots\rangle+\tilde{c}_{2}|1;010\ldots\rangle+\tilde{c}_{3}|1;001\ldots\rangle +\ldots \nonumber \\
&& +c_{0}|0;000\ldots\rangle ,\label{eq:matrixstate3}
\end{eqnarray}
where $|c_{0}|^{2}=\sin^{2}\alpha$ and $\sum_{j}|\tilde{c}_{j}|^{2}=\cos^{2}\alpha$.
Tracing over all qubits of $B$ except $B_{1}$ gives
\begin{eqnarray}
\rho_{AB_{1}} & = & \bigl\{\tilde{c}_{1}|1;1\ldots\rangle+c_{0}|0;0\ldots\rangle\bigr\} \times\bigl\{\tilde{c}_{1}^{*}\langle1;1\ldots|+c_{0}^{*}\langle0;0\ldots|\bigr\} \nonumber \\
 && +\bigl\{|\tilde{c}_{2}|^{2}+|\tilde{c}_{3}|^{2}+\ldots\bigr\}|1;0\ldots\rangle\langle1;0\ldots| ,\label{eq:redrho1}
\end{eqnarray}
which is of the $X$-form (\ref{eq:matrixrho}), from which the concurrence
can be calculated. More generally, we find $C_{AB_{J}}=2|\tilde{c}_{J}||c_{0}|$,
to confirm the required result.

We can see directly from (\ref{eq:fourqubitstate}) that all nonlocal pairwise entanglement $C_{IJ}$ can vanish (SSPC$=0$), even where maximal global entanglement $C_{AB}$ exists. This is consistent with behavior of other multi-partite states, such as the GHZ state
$(|000\rangle+|111\rangle)/\sqrt{2}$, which display no
bipartite entanglement once one of the qubits is traced out~\cite{vidthreequbit}.
Calculation reveals the four qubit state~(\ref{eq:matrixstate3})
has zero bipartite concurrence for all nonlocal bi-partitions when there are equal probability amplitudes: $|c_{11}|=|c_{12}|=|c_{21}|=|c_{22}|=|c_{0}|=1/\sqrt{5}$. 

We note however this loss of entanglement is not the case for all nonlocal partitions. The Theorem 2 reveals for $|\Phi\rangle$ states that entanglement $C_{AJ}$ defined for nonlocal partitions $\left\{ a_{1},\ldots,a_{N}\right\} \left\{ b_{J}\right\} $ satisfies a simple addition rule~(\ref{eq:addipsi}), $\sum_{J}C_{AJ}^{2}=C_{AB}^{2}$. Thus for the case of $4$ qubits, 
\begin{equation}
C_{\{1,2\}\{1,2\}}^{2}=C_{\{1,2\}\{1\}}^{2}+C_{\{1,2\}\{2\}}^{2} ,\label{eq:rulesum3-1}
\end{equation}
which implies a zero $3$-tangle between $A$ and qubits $b_{1}$ and $b_{2}$ in this case, as for the $W$-state considered above. However, there is a nonzero $3$-tangle for systems $A_{1}, B_{1}, B_{2}$, for example, defined as $\tau=C_{A_{1}B_{1}B_{2}}^{2}=C_{A_{1}B}^{2}-C_{A_{1}B_{1}}^{2}-C_{A_{1}B_{2}}^{2}$. For the state, $C_{A_{1}B}$ can be calculated similarly to $C_{AB_{1}}$ (see~(\ref{eq:matrixstate3}) to yield a nonzero value, which implies
a $3$-tangle of $\tau_{3}=C_{A_{1}B}^{2}$ when $C_{A_{1}B_{1}}=C_{A_{1}B_{2}}=0$,
that is, when the probability amplitudes are equal. This gives a different
behavior to the $4$-qubit GHZ-type states, which have zero $3$-tangle
upon tracing out over one of the four parties~\cite{fourqubit}. Such GHZ states cannot be formed from the two-qubit GHZ-type state $|\Phi\rangle$ under the local transformations considered.

\section{Jaynes-Cummings example}

We give an example of the conservation rule by examining the case of $N,M=2$, where the qubits $a_{1}, b_{1}$ are two-level atoms with transition frequency $\omega_{0}$, and the
qubits $a_{2}, b_{2}$ are cavity modes with resonant frequencies $\omega_{a}$ and $\omega_{b}$, respectively. We consider the local interaction Hamiltonian of the Jaynes-Cummings form~\cite{jc}: 
\begin{equation}
H_{A}=\hbar\omega_{0}S_{A}^{z}+\hbar\omega_{a}\left(a^{\dagger}a+\frac{1}{2}\right)+\hbar g_{a}\left(a^{\dagger}S_{A}^{-}+aS_{A}^{\dagger}\right) ,\label{e2}
\end{equation}
where $S_{A}^{+}$, $S_{A}^{-}$ and $S_{A}^{z}$ are respectively raising, lowering
and spin-$z$ operators for the atom qubit $a_{1}$, and $a^{\dag}\ (a)$
are the creation (annihilation) operators for the cavity mode qubit
$a_{2}$. The parameter $g_{a}$ is the strength of the coupling between the
atom and the cavity mode. The local Hamiltonian for $B$ is defined similarly.

The Jaynes-Cummings interaction is fundamental in describing couplings
between field and atoms in cavities~\cite{cavityqed}. The Jaynes-Cummings
interaction Hamiltonian has also been used to model the qubit-cavity
coupling in circuit QED experiments that use superconducting qubits,
and which have recently realized an entangled two-qubit nonlocal Bell
state~\cite{circuitqed,mc07}. Our conservation results enable prediction
of the entanglement between cavity-atom pairs, for the two types of
Bell-state entanglement, and could in principle be tested in these
experimental situations as well as in the all-optical entanglement
experiment of Almeida \etal~\cite{alexp}.
\begin{figure}[h]
\begin{center}
\includegraphics[clip,width=0.6\columnwidth]{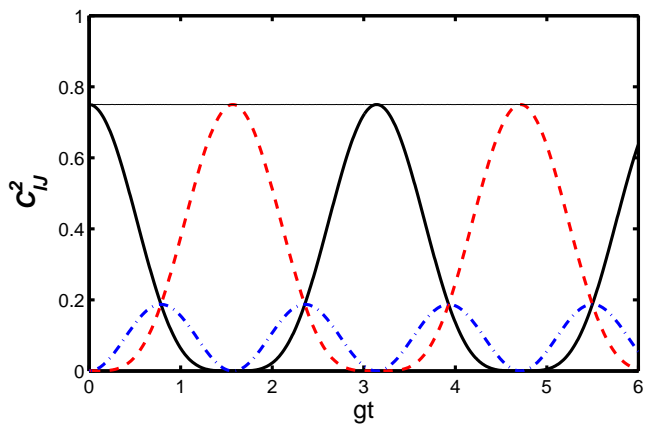}
\end{center} 
\caption{Evolution of the two-qubit concurrences for state $|\Psi\rangle$ with the Hamiltonian (\ref{e2}) and for different nonlocal partitions: $C_{11}^{2}$ (solid line); $C_{22}^{2}$ (dashed line); $C_{12}^{2}=C_{21}^{2}$ (dash-dotted line); $C_{AB}^{2}$ (thin solid line). The following
rule, universal for $|\Psi\rangle$, holds: SSPC$=C_{AB}^{2}$. Here
$\Delta_{a}=\Delta_{b}=0$, $g_{a}=g_{b}$ and $\alpha=\pi/6$.
In the case of symmetric interactions $g_{a}=g_{b}$ illustrated here, the simple
conservation rule $C_{AB}=C_{11}+C_{22}$ of Yonac \etal~\cite{yonac,yonac1} also holds.}
\label{cap:conservlawsingle2-1}
\end{figure}

\subsection{Jaynes-Cummings example for $|\Psi\rangle$ }

\begin{figure}[h]
\begin{center}
\includegraphics[clip,width=0.6\columnwidth]{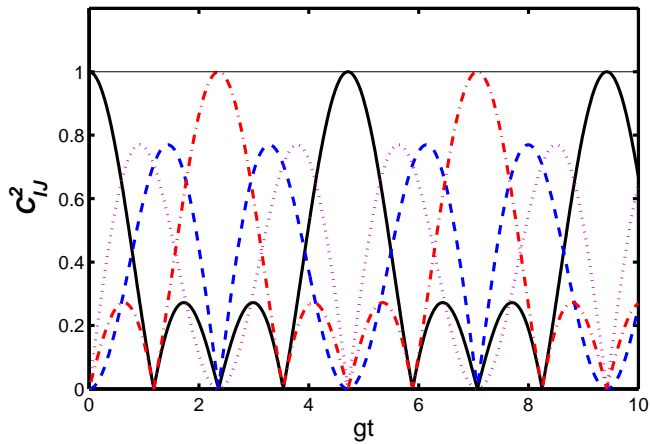}
\end{center} 
\caption{Evolution of two-qubit concurrences for state $|\Psi\rangle$ with Hamiltonian~(\ref{e2}) and for different nonlocal partitions. The plots confirm the rule, universal for $|\Psi\rangle$, that SSPC$=C_{AB}^{2}$. $C_{11}$ (solid line), $C_{22}$ (dashed line), $C_{12}$ (dashed-dotted line), $C_{21}$ (dotted line). Here, $\Delta_{a}=\Delta_{b}=0, g_{a}=2g_{b}, g=(g_{a}+g_{b})/2$, and 
$\alpha=\pi/4$.}
\label{cap:conservlawsdiffg-1}
\end{figure}

Solutions for the concurrences describing pairwise entanglement can be evaluated for the case of an initial Bell state entanglement of type $|\Psi\rangle$, for the full case of arbitrary couplings and
detunings, defined as $\Delta_{a}=(\omega_{0}-\omega_{a})/2$ and $\Delta_{b}=(\omega_{0}-\omega_{b})/2$. Figures~\ref{cap:conservlawsingle2-1} and \ref{cap:conservlawsdiffg-1} show concurrences for symmetric and asymmetric interactions, to confirm the conservation law~(\ref{eq:concurrenceequal}). 

\subsection{Jaynes-Cummings example for $|\Phi\rangle$}

\begin{figure}[h]
\begin{center}
\includegraphics[clip,width=0.6\columnwidth]{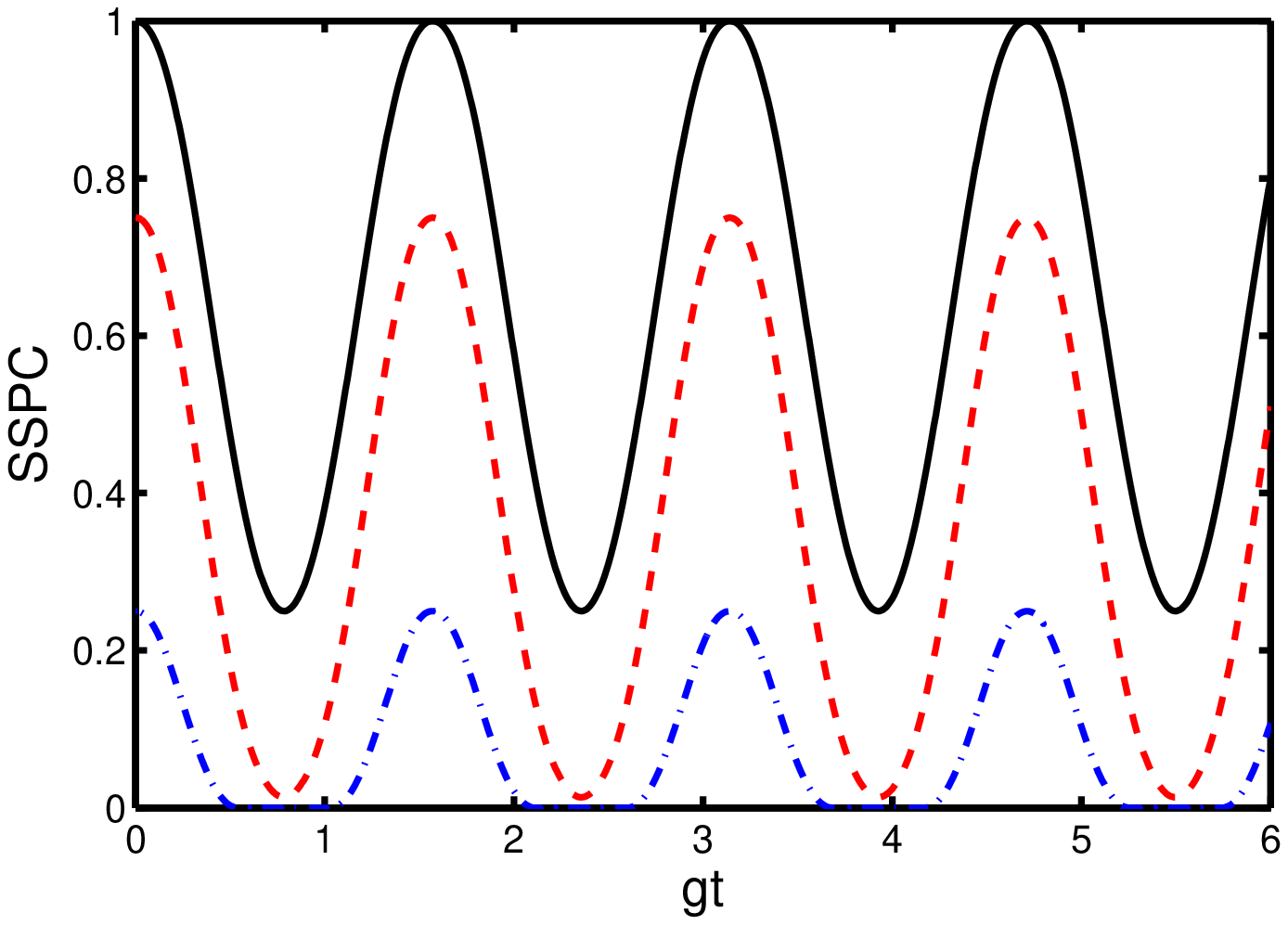}
\includegraphics[width=0.6\columnwidth]{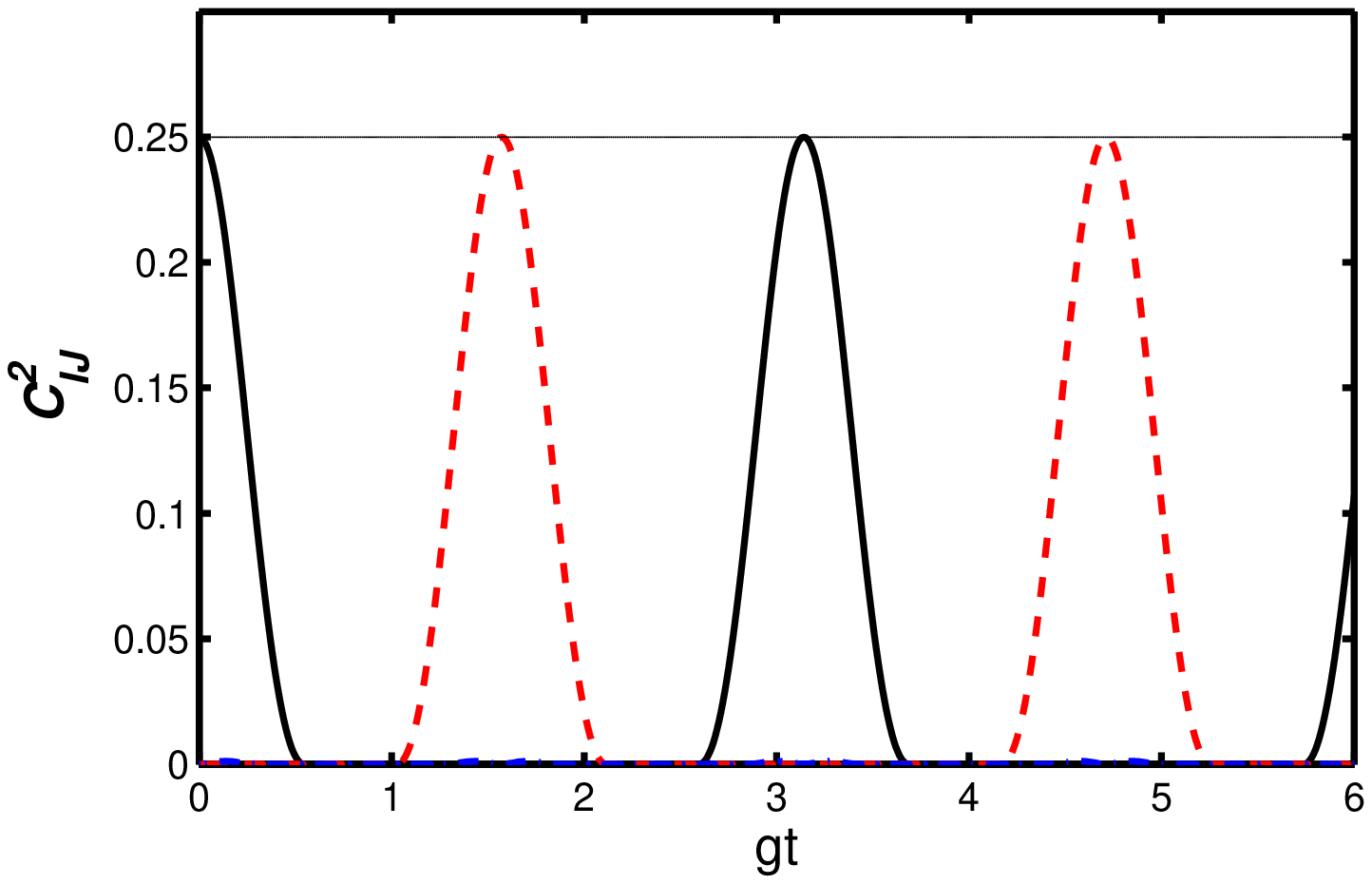}
\end{center}
\caption{Evolution of pairwise concurrences for different non-local partitions, for state $|\Phi\rangle$ and Hamiltonian~(\ref{e2}). Here, $\Delta_{a}=\Delta_{b}=0$ and $g_{a}=g_{b}=g$. The top figure plots SSPC for $\alpha=\pi/4$ (solid line), $\alpha=\pi/6$ (dashed line), $\alpha=\pi/12$ (dashed-dotted line). The bottom figure plots the individual concurrences for $\alpha=\pi/12$: $C_{11}^{2}$ (solid line), $C_{22}^{2}$ (dashed line), $C_{AB}^{2}$ (thin solid line). $C_{12}^{2}=C_{21}^{2}$ (dash-dotted line) is below $0.01$ in this case. As the global entanglement
$C_{AB}$ weakens, regions of total loss of nonlocal pairwise entanglement
are evident (SSPC=0).} 
\label{cap:pairwiseloss9ab}
\end{figure}

The loss of all nonlocal pairwise concurrence for evolution of the Bell state $|\Phi\rangle$ is evident in the model~(\ref{e2}). This is illustrated in figure~\ref{cap:pairwiseloss9ab}, where we plot the time evolution of pairwise concurrences for different non-local partitions and for state $|\Phi\rangle$.  Where the entanglement $C_{AB}$ is low,
we can identify regions where each $C_{IJ}$ is zero (SSPC$=0$).
The SSPC is regained when the transfer of entanglement from qubits
$a_{1},b_{1}$ into the qubits $a_{2},b_{2}$ is complete, so that
the inequality of Theorem 2 is saturated at regular intervals determined
by the Rabi frequency, at which all excitations (qubits with bit value
$1$) are in the same type of qubit (atoms or field modes). We see that $C_{AB}$ is low and observe the feature predicted by Lopez \etal~\cite{lopez} for reservoir interactions that the {}``birth'' of entanglement in cavity modes is delayed a finite time after the {}``death'' of entanglement in the atoms.

\section{Effect of environment on entanglement transfer}

\begin{figure}[h]
\begin{center}
\includegraphics[clip,width=0.6\columnwidth]{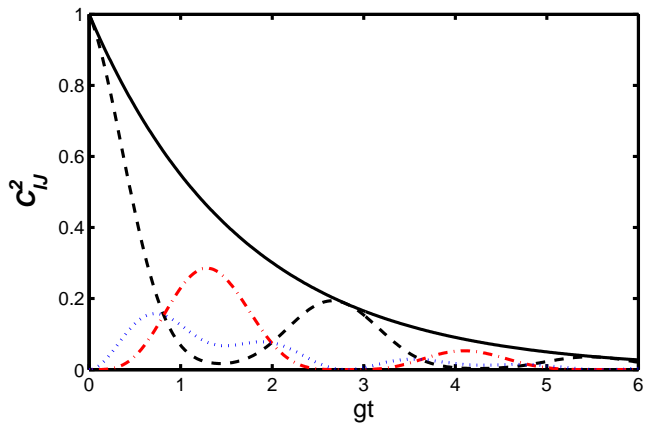} 
\end{center}
\caption{Evolution of pairwise concurrences for the state $|\Psi\rangle$ under the Hamiltonian~(\ref{e2}) and with coupling to reservoirs. $C_{11}^{2}$ (dashed line); $C_{22}^{2}$ (dashed-dotted line); $C_{12}^{2}=C_{21}^{2}$ (dotted line); $C_{AB}^{2}$ (solid line). Here, $\Delta_{a}=\Delta_{b}=2g$, $g_{a}=g_{b}=g$, $\alpha=\pi/4$, and $\gamma=0.3g$.}
\label{lfig5}
\end{figure}

We show that the constituent addition rule~(\ref{eq:concurrenceequal})
for Bell state $|\Psi\rangle$ also applies to describe entanglement
transfer for open systems, where energy loss is modeled by zero-temperature
reservoir interactions
\begin{equation}
H_{R}=a_{I}\Gamma^{\dagger}+a_{I}^{\dagger}\Gamma ,\label{eq:res}
\end{equation}
if the qubits are cavity modes, or 
\begin{equation}
H_{R}=\sigma_{I}\Gamma^{\dagger}+\sigma_{I}^{\dagger}\Gamma ,
\end{equation}
if the qubits are two-level atoms. Here, $\Gamma=\sum_{R}g_{R}b_{R}$
and $b_{R}$ is the boson destruction operator for one of many vacuum
modes that comprise the reservoir. Inclusion in the Hamiltonian~(\ref{e2}) of such coupling, under the Markovian assumption and assuming equal cavity and atomic damping rates $\gamma$, leads to a simple damping envelope for the concurrences $C_{IJ}(t)\rightarrow C_{IJ}(t)\exp(-\gamma t)$, as illustrated in figure~\ref{lfig5}. The global entanglement $C_{AB}$ shows asymptotic decay $C_{AB}(t)=C_{AB}(0)\exp(-\kappa t)$, where $\kappa$ is the global decay rate, to confirm the rule $\sum_{IJ}C_{IJ}^{2}(t)=C_{AB}^{2}$ in this case. Finally, we point out the conservation rule will apply to other qubit-number conserving interactions, and can be studied in relation to other models of decoherence such as due to dephasing, important to, for example, electron spins in quantum dots~\cite{los,yk08}.

While the one-sided addition rule (\ref{eq:addipsi}) holds for both types of Bell state entanglement, the different nonlocal pairwise concurrence relations allow us to deduce that the robustness of this addition rule with respect to coupling to the environment will be very different. If the system $a_{1}$ initially prepared in a $|\Phi\rangle$ state is coupled to an environment
via interactions such as~(\ref{eq:res}), the inequality indicates that there can be a complete loss of pairwise concurrence between $a_{1}$ and $b_{1}$ at finite times (ESD)~\cite{yueberly}.

\section{Conclusion}

How entanglement is lost, or transferred, when a system is coupled
to an environment is a fundamental issue. In this paper we have quantified
how such transfer takes place in a useful subset of scenarios, where
the initial entanglement is in the form of a two-qubit Bell state,
and the local interactions preserve total qubit number at each site.
When the Bell state has only one excitation, there is a conservation
of the sum of the tangle (the concurrence squared) associated with
each resulting non-local two qubit bipartition. When the Bell state
is a superposition of zero and two excitations, the sum of the tangle
for the non-local bipartite partitions has an upper bound, but can
be zero, to indicate a total absence of pairwise two qubit entanglement.
We have shown that a conservation rule for a nonlocal partition
can be found in both cases. This is that the sum of the tangle
between qubit $A$ and each of the qubits at $B$ is conserved
throughout the evolution. In order to further analyze the entanglement
distributed among the qubits, new analyses of measures for multipartite
entanglement could be useful~\cite{hiesmulti}.

The rules derived in this paper apply where qubit number at each location
is conserved, and hence are fundamental to a quantitative understanding
of the distribution of the entanglement that will be inevitably shared
between a system and its environment, because of decoherence mechanisms.
Our results should be directly testable in the experimental arrangements
such as those of Almeida \etal~\cite{alexp}.

\ack
This work was funded by the Australian Research Council and the ARC
Center of Excellence program.

\end{document}